\documentclass[preprint,12pt]{elsarticle}
\usepackage{amssymb}
\usepackage{url}
\usepackage{graphicx}
\usepackage[usenames,dvipsnames]{xcolor}
\usepackage[section]{placeins}
\usepackage[normalem]{ulem}
\usepackage{xcolor}
\journal{Physics Letters B}

\begin{document}

\begin{frontmatter}
\title{\textit{Ab initio} calculations of $^5$H resonant states}
\author{R. Lazauskas} \address{IPHC, IN2P3-CNRS/Universit\'e Louis Pasteur BP 28, F-67037 Strasbourg Cedex 2, France}
\author{E. Hiyama} \address{Department of Physics,  Kyushu University, Fukuoka, Japan, 819-0395 and \\  RIKEN Nishina Center, 2-1 Hirosawa, Wako 351-0106, Japan}
\author{J.~Carbonell}     \address{Institut de Physique Nucl\'eaire, Universit\'e Paris-Sud, IN2P3-CNRS, 91406 Orsay Cedex, France}

\begin{abstract}
By solving the 5-body Faddeev-Yakubovsky equations in configuration
space with  realistic nuclear Hamiltonians we have studied the
resonant states of $^5$H isotope. Two different methods, allowing
to bypass the exponentially diverging boundary conditions,
have been employed providing consistent results.
The existence of $^5$H broad J$^{\pi}$=1/2$^+$,3/2$^+$,5/2$^+$  states as S-matrix poles has been confirmed
and compared with the, also calculated,  resonant states in $^4$H isotope.
We have established that the positions of these resonances only mildly depend on the nuclear interaction model.
 \end{abstract}


\end{frontmatter}

\section{Introduction} \label{intro}

Neutron-proton unbalanced nuclei are a unique laboratory to study
the isospin dependence of the nuclear interaction as well as the
nuclear properties far beyond the stability line. Hydrogen
isotopes play a special role, as they provide the extreme neutron
to proton ratios and thus are actively studied both theoretically as well as experimentally.

\bigskip
 If there seems to be a general agreement between the
experimentalist on the existence of several resonances  in $^5$H isotope
\cite{Young_PR173_1968,Korsheninnikov_PRL87_2001,
Golovkov_et_al_2003_04_05,Wuosmaa_PRC95_2017}, the
parameters of these resonances vary significantly from one
experiment to another. These discrepancies might be
attributed to the different reaction mechanism leading to the
formation of  broad resonant states and to the difficulty in
identifying their quantum numbers. Indeed, in any physical
observable, the contributions  of broad resonances strongly
interfere among themselves as well as with a non-resonant
background, producing in the resonance region a response, which strongly depends on the particular reaction mechanism.
Unless the reaction mechanism is well understood
and one is able to isolate -- analytically or numerically -- both contributions, the resonance parameters are not defined in a unique way.
In this case only qualitative analysis of the experimentally extracted broad resonance parameters makes sense.

According to the results of Korsheninnikov et
al.~\cite{Korsheninnikov_PRL87_2001}  the lowest resonant state of
$^5$H would be narrower  than the ones of $^4$H isotope, a fact that
could be naively  attributed to the additional attraction provided
by the $4n$ subsystem present in $^5$H relative to $3n$ one
present in $^4$H. Or, alternatively, to the $nn$ pairing effect over a core of $^3$H. By extending this effect to $6n$ one could in
this way  obtain  a natural explanation for the extremely narrow
parameters of the $^7$H resonance  published in~\cite{H7_GANIL_2007}.

The findings of ref.~\cite{Korsheninnikov_PRL87_2001}  were
confirmed by Golovkov et al.~\cite{Golovkov_et_al_2003_04_05},
attributing even smaller widths to the $^5$H resonant states. They are
however in conflict with a recent work~\cite{Wuosmaa_PRC95_2017}
where the resonance width was found to be much larger and
comparable with the one of the $^4$H ground state, although this
difference could be partially attributed to the different reaction
mechanism used in their production.

\bigskip
Several theoretical works  on this system were performed in the
past. They could be classified in two major groups:  t-n-n models in which $^3$H nuclei is considered as a
point like particle (hereafter denoted  by $t$) and microscopic
3-body cluster models, based on $^3$H+n+n structure. In t-n-n
models, the Pauli principle between the valence neutrons and those
contained in the $^3$H core is mimicked by a repulsive term in the effective t-n interaction.
In contrast, microscopic models start
from a bare Nucleon-Nucleon (NN) interaction and employ properly antisymmetrized
5-nucleon wave functions, however some severe constrains on the
dynamics of the $^3$H core are applied.

These studies were pioneered by Shul'gina
et al.~\cite{SDGB_PRC62_2000}, where $^5$H is described by a t-n-n
model, solved by means of hyperspherical Harmonics techniques. The
t-n potential was taken from the same authors
\cite{Grigorenko_nt_PRC_1999} and the $nn$ one from Gogny et al.
\cite{GPT_nn_PLB32_1970}.
Within the same approach, a theoretical study of broad states in
few-body systems beyond the neutron drip line was published
in~\cite{GTZ_EPJA_2004}. No precise positions were given but
studies of the sensitivity of the $^5$H spectrum to reaction
mechanism predict that the "ground state" peak may be observed at
2-3 MeV as lowest position, though could be shifted to much higher
energies in certain situations.

In~\cite{DGFJ_NPA786_2007} a similar 3-body model but with a
refined t-n interaction  was considered with $V_{nn}$ taken
from~\cite{V_nn_GFJ_PRC69_2004}. The positions of $^5$H  resonant
states were computed  by means of the Complex Scaled (CS)
hyperspherical adiabatic expansion method. However, this model
failed to obtain realistic $^5$H resonant states when using only
two-body interactions;  adjustment of an additional {\it ad-hoc}  t-n-n
3-body force was required  to reproduce the positions of the
J$^\pi$=1/2$^+$,3/2$^+$,5/2$^+$ states. The energy distributions
of  the fragments after their decay were computed as well.

A t-n-n model was also used,  in  the framework of hypernuclear
physics~\cite{HOKY_NPA908_2013}, to compute $^5$H resonant states.
The  problem was solved by combining CS with  the Gaussian
expansion method~\cite{HOKY_NPA908_2013}.
The ${nn}$ interaction was  Argonne V8'~\cite{AV18_1995}  and  $t-n$ one was adapted  from~\cite{SDEBZ_NPA597_1996}. 
It was found again, that this 3-body model failed to reproduce ground state resonance parameters unless some ad-hoc 3-body force was included and properly adjusted.

\bigskip
The first microscopic model for $^5$H using the Generator
Coordinate Method (GCM) was used  in~\cite{DK_PRC63_2001}. The key
ingredient of the calculation, the NN potential, was based on
semi-realistic Minnesota (MN) \cite{MN_VNN_TLY_NPA286_1977}    and
Mertelmeier and Hofmann \cite{MH_VNN_NPA459_1986} models.
Resonance parameters were identified using the analytic
continuation in a coupling constant (ACCC)
method~\cite{ACCC_Kukulin_1989}. In a later
work~\cite{AD_NPA813_2008} this last study was improved, replacing
the GCM technique by  a 3-body microscopic approach based on
Hypersherical Harmonics method.

A microscopic three cluster model of $^5$H, using  the equivalent Resonating Group Method (RGM) approach and
 the  MN potential, was also presented in~\cite{Arai_PRC66_2003}.

Finally,   a microscopic cluster model of $^5$H  was presented
in~\cite{ABV_JPCS_2008}, based on 3-cluster J-matrix method
(MJM) and Hyperspherical Harmonics basis. Resonance positions were
directly computed by imposing the appropriate boundary conditions.
Different components of the MN and modified Hasegawa-Nagata
potentials \cite{Hasegawa_Nagata} were used to built the NN interaction.

\bigskip
It seems firmly established that the t-n-n models of $^5$H
systematically require a properly adjusted t-n-n 3-body force in order to
produce $^5$H resonant states with reasonable widths. Thus, their
predictions critically depends on  3-body force parameters on which there is
little control. On the contrary, all microscopic models agree on the
existence of a relatively broad J$^\pi$=1/2$^+$ resonant state in
$^5$H, which is located $\approx$ 1.5 MeV above the $^3$H
threshold and has a width in the range of $\Gamma \approx$ 1-3
MeV. The presence of even broader J$^\pi$=3/2$^+$ and J$^\pi$=5/2$^+$
states has been also indicated. These methods however rely on
3-cluster wave functions and employ simplistic phenomenological NN
interactions. As we will see in what follows, their predictions
display also some dispersion even when using the same interaction.

We believe it could be of interest to determine unambiguously
the S-matrix pole position of these resonant states by undertaking
the {\it ab initio} solution of this problem with the only input of a realistic NN interaction. Indeed, our recently developed
techniques, allowing to solve the 5-body Faddeev-Yakubovsky equations
in configuration space, and for the first time applied to
calculate n-$^4$He elastic phase shifts \cite{Rimas_5N_FBS_2018,Rimas_5N_PRC97_2018}, provide us the
opportunity to investigate the properties of $^5$H nucleus from an
{\it ab initio} theoretical perspective. By solving the 5-nucleon
problem in the complex energy plane we present here the first {\it
ab initio} study  of $^5$H, involving no approximations on the dynamics.

A new experiment has been recently performed at RIKEN \cite{Miguel_RIKEN_NP1512}, combining
benefits of radioactive ion beam facilities and multineutron
detection arrays, aiming to resolve the puzzle of observable
$^4$n and $^7$H  states.
Understanding the evolution of H-isotope resonances, {originated from S-matrix poles in the vicinity of the real energy axis},
when increasing the number of neutrons
 was  one of the main
motivations of our work. Therefore our theoretical attempt is
timely and hopefully will also be relevant for a better understanding and interpretation of the experimental data.

\bigskip
The paper is organized as follows. In Section \ref{Formalism} we
summarize the formalism used to solve the 5-body problem and
the extrapolation to the complex energies. Numerical results for
$^4$H and $^5$H resonant states are given in Section
\ref{Results}. Section \ref{Conclusion} contains some concluding
remarks.

\section{Formalism}\label{Formalism}

Faddeev-Yakubovsky (FY) equations were formulated
in~\cite{Yakubovsky_SJNP5_1967}, as an extension of the seminal
work  of Faddeev~\cite{Fad_JETP39_1960}  for  a 3-particle system (A=3) to an arbitrary number of particles (A=N).
Starting from the middle 80's, they have been successfully solved both in momentum and configuration space for  A=4 in all
possible bound and scattering channels  \cite{Nogga_4a,LazCarb_inoy,DeltFons_4N}.
The first results for A=5 problem, recently obtained by one of the authors  \cite{Rimas_5N_FBS_2018,
Rimas_5N_PRC97_2018}, introduced  the formalism  in some detail. In this letter we will  just highlight the main ideas.

\bigskip
\begin{figure}[h!]
\centering\includegraphics[width=16.cm]{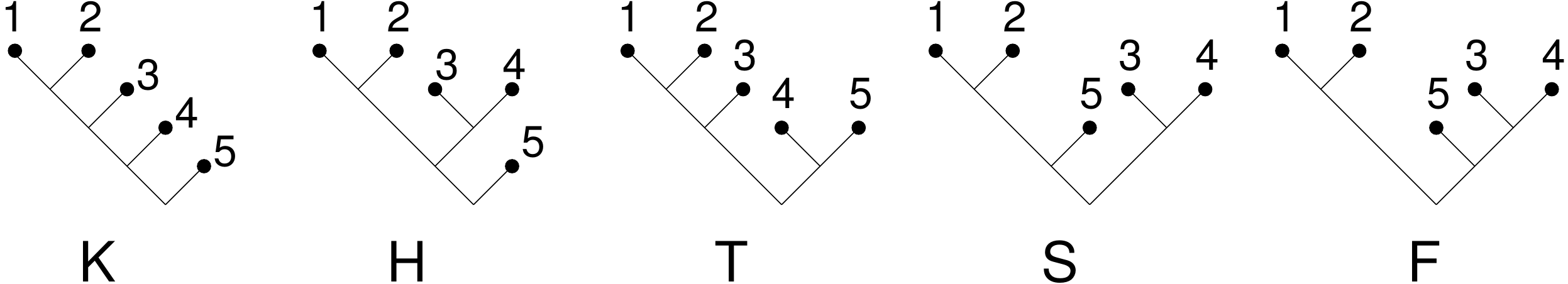}
\vspace{0.5cm}

\centering\includegraphics[width=16.cm]{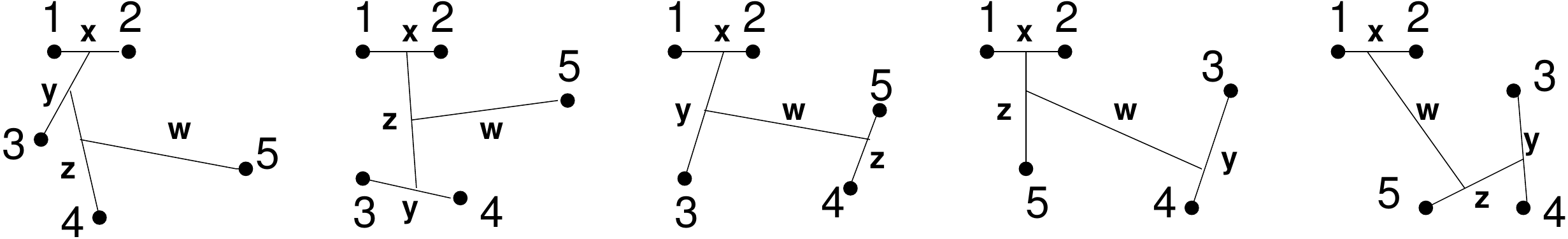}
\caption{Different topologies for the 5-body FY amplitudes and corresponding Jacobi coordinates.}\label{Trees}
\vspace{0.cm}
\end{figure}

In the FY approach, the intrinsic (center of mass) 5-body wave
function depending on four vector variables $\Psi(\vec{x},\vec{y
},\vec{z},\vec{w})$,  is decomposed into 180 components
$F^{(i)}_a$  corresponding to the five different topologies in the
cluster partition of the system, which are represented in Fig.
\ref{Trees}. Subscript $a=K,H,S,T,F$ denotes their topological
type and  the superscript $i$ the sequence of particle indexes.

These components are naturally expressed in their proper set of
Jacobi coordinates $\vec{x}^{(i)}_a$, $\vec{y}^{(i)}_a$, $\vec{z}^{(i)}_a,\vec{w}^{(i)}_a$.
These are represented  in the lower panel of Fig.\ref{Trees} and have the generic form
\[  (\vec{x},\vec{y},\vec{z},\vec{w} ) = \sqrt{ 2\mu_{st}\over m_0} \left( \vec{R}_s- \vec{R}_t  \right)\]
where $\vec{R}_i$ denotes the center of mass positions of the two clusters (eventually single particles) connected by each coordinate
and $\mu_{st}$ their reduced mass.  An arbitrary mass $m_0$ is introduced to keep the physical units.

For  a system of identical particles with mass $m$ one chooses
$m_0=m$. One can also choose a preferential particle ordering $i$,
say $i=\{1,2,...,5\}$, and the permutation symmetry allows to
reduce the number of independent components to only five $\{F_a\}$,
one for each topological type. The remaining components, as well
as the total wave function, can be reconstructed from this subset
with the  help of coordinate basis transformation operators $\hat{P}^{b,i}_a$:
\begin{eqnarray}
F_a^{(i)}  (\vec{x}^{(i)}_b,\vec{y}^{(i)}_b,\vec{z}^{(i)}_b,\vec{w}^{(i)}_b)= \hat{P}^{b,i}_a F_a(\vec{x}_a,\vec{y
}_a,\vec{z}_a,\vec{w}_a); \label{eq:PPaaa} \\
 \Psi(\vec{x}^{(i)}_b,\vec{y}^{(i)}_b,\vec{z}^{(i)}_b,\vec{w}^{(i)}_b)=\sum_{j}\sum_{a=1}^5  F^{(j)}_a(\vec{x}^{(i)}_b,\vec{y}^{(i)}_b,\vec{z}^{(i)}_b,\vec{w}^{(i)}_b).
\end{eqnarray}
Using the  completeness relation one may split any
operator $\hat{P}^{b,i}_a$  into a product of simpler
transformation operators $\hat{P}_k$ , similar to the ones used in the 3-body case \cite{Rimas_5N_PRC97_2018}, which affect only two vector coordinates at once and
let unchanged the functional dependence of the remaining  two:
\begin{eqnarray}
\hat{P}^{b,i}_a ={\prod}_{k} \hat{P}_k.
 \label{eq:PPccc}
 \end{eqnarray}
As discussed in~\cite{Rimas_5N_PRC97_2018} a set of eleven such
operators is sufficient to express any complicated transformation appearing in the solution of a 5-body problem.

The FY components $F^{(i)}_a$ correspond to the different
asymptotic channels and satisfy a coupled system of second order partial differential equations.
In view of solving these equations,  a partial wave expansion on each vector variables is performed
\[ F_a ( \vec{x}_a,\vec{y }_a,\vec{z}_a,\vec{w}_a)=   \sum_{\alpha} {1\over x_ay_az_aw_a}  \; f_{a,\alpha}(  x_a,y_a,z_a,w_a  ) \;  \mathcal{Y}_{\alpha} (\hat{x}, \hat{y},\hat{z},\hat{w})  \]
where  $\mathcal{Y}_{\alpha}$ is a generalized "quadripolar
harmonic" function accounting for the angular momentum, spin and isospin couplings
\[  \mathcal{Y}_{\alpha} =  \left[    \left[   [ l_x, l_y ]_{l_{xy}}  \; [l_z,l_w]_{l_{zw}} \right]_L  \; \{S\}_S    \right]_{J} \{T\}_{TT_z} ,  \]
$\{S \}_S\!=\mid \![[s_1,s_2]_{s_x}  [s_3,s_4]_{s_y}]_{s_{xy}} s_5
;SS_z\rangle$   ($\{T \}_{TT_z} \!= \mid \! [[t_1,t_2]_{t_x}
[t_3,t_4]_{t_y}]_{t_{xy}} t_5 ;TT_z\rangle$) holds for the spin
(isospin) couplings and $\alpha$ labels the set of quantum numbers
involved in the intermediate couplings. The above expressions
represents the main coupling scheme used in this work for the  K, H, S and F topologies. For T topology
a slightly modified coupling scheme was used, obtained by permuting particle 3 and 5 indexes for spin and isospin couplings.
These coupling schemes allow to (anti)symmetrize the total wave function in a trivial way.

After projecting the angular part, one is left with a set of
coupled four-dimensional integro-differential equations for the reduced radial amplitudes $f_{a,\alpha}(x,y,z,w)$ in the form
\begin{small}
\begin{equation}\label{EFY5}
 \left[   q^2\delta_{\alpha\alpha'} + \Delta^{\alpha}_{xyzw}\delta_{\alpha\alpha'} - v_{\alpha\alpha'}(x)  \right]  f_{a\alpha'}(x,y,z,w)= v_{\alpha\alpha'}(x)  \sum_{b\beta} \int d\theta d\xi d\zeta \;  h^{b\beta}_{a\alpha'}(x,y,z,w,\theta,\xi,\zeta)   f_{b\beta}(x_b,y_b,z_b,w_b)
\end{equation}
\end{small}
where
\[\frac{m}{\hbar^2}  \Delta^{\alpha}_{\alpha,xyzw} = \partial^2_x+ \partial^2_y+\partial^2_z+\partial^2_w  - {l_x(l_x+1)\over x^2}-  {l_y(l_y+1)\over y^2} - {l_z(l_z+1)\over z^2} - {l_w(l_w+1)\over w^2} \]
and $v_{\alpha\alpha'}$ is the two-body potential (multiplied by $\frac{m}{\hbar^2}$).

In the right hand side of  eq. (\ref{EFY5}), an integration over
the angular variables $\theta,\xi$ and $\zeta$, allowing to
connect the Jacobi coordinate sets $(x,y,z,w)$ and
$(x_b,y_b,z_b,w_b)$, is performed. This essential operation is
realized in the matrix form by breaking the integration in three
successive steps,  as briefly explained in
eqs.(\ref{eq:PPaaa}-\ref{eq:PPccc}). The radial dependence of
these amplitudes is sought by expanding them in  the Lagrange-Laguerre
basis functions using the Lagrange-mesh method~\cite{Baye_PR565_2015}.
In our numerical calculation, the partial wave basis was limited to
${\rm max}(l_x,l_y,l_z,l_w)<4$. 
The number of Lagrange-mesh points to expand the radial FY components  was 
$n_x=12-2l_x$  for the $x$-coordinate and $n_\alpha=11-2l_\alpha$  for $y,z$ and $w$. 
The errors introduced by these truncations are smaller than the uncertainties related with the methods
used to calculate the positions of the resonant states.

\bigskip
To access the complex energy eigenvalues of a resonant state, we
have used the method of Analytic Continuation in the Coupling
Constant (ACCC) in the same form as in our previous works
\cite{LC_3n_PRC72_2005,LC_4n_PRC72_2005,HLCK_PRC93_2016}  and we
have developed a variant of the Smooth Exterior Complex Scaling
method (SECS)  that will be detailed in what follows.

ACCC does not allow a direct determination of  resonance parameters.
These are  obtained by  artificially binding the system with an ''external field''
 driven by a  strength parameter  $\lambda$
and computing its binding energy  as a function of this parameter:
$B(\lambda)$. In our case, we added to $^5$H Hamiltonian a
fictitious 5-body hyperradial potential on the form:
\begin{equation}
V_{5}(\rho)= -\lambda \; \rho^p \; e^{-\rho^2/\rho^2_0} \qquad \rho=\sqrt{x^2+y^2+z^2+w^2}   \label{eq:accc_pot}
\end{equation}
Equation (\ref{EFY5}) should be accordingly modified by
adding to its left hand side the term
$V_{5}\delta_{\alpha\alpha'}f_{a\alpha'}$.

Several values of  $B_{^5H}(\lambda)$ were calculated  up to $^3$H+n+n threshold, corresponding to the critical strength $B_{^5H}(\lambda_0)\equiv
B_{^3H}$, above which the $^5$H S-matrix pole moves into the resonance region.
These values are used to determine a Pad\'{e}-like  functional dependence in the domain of real arguments
\[  \sqrt{ B_{^5H}(\lambda)-B_{^3H} }= f_{ACCC}( \sqrt{\lambda-\lambda_0}  )  \]
which  will be analytically continued in the complex plane until  $f_{ACCC}(\sqrt{-\lambda_0})$,  where the external potential $V_{5}$
is fully removed ($\lambda=0$). At this point, the  position of the $^5$H
resonance relative to $^3$H threshold may be easily determined. As
it was explained in \cite{HLCK_PRC93_2016,CLHK_FBS_2017,LHC_PTEP7_2017}, in the context of 3n and 4n
systems, the use of an artificial 5-body force  to bind $^5$H is motivated by the compelling need to avoid spurious branching
points, related to the appearance of artificially bound $^5$H subsystems.

\bigskip
In our previous studies we have extensively employed the CS method to
calculate resonance parameters as well as several scattering
observables \cite{CL_CS_SCAT_2011_13}. Nevertheless, the conventional CS fails in
describing broad resonant states for they require a large rotation angle $\theta$ to be exposed.
Furthermore, some short range potentials, like e.g. chiral NN interactions, include sharp vertex form factors and
start to diverge already when very moderate values of $\theta$ are used.

\begin{figure}[h!]
\centering\includegraphics[width=9cm]{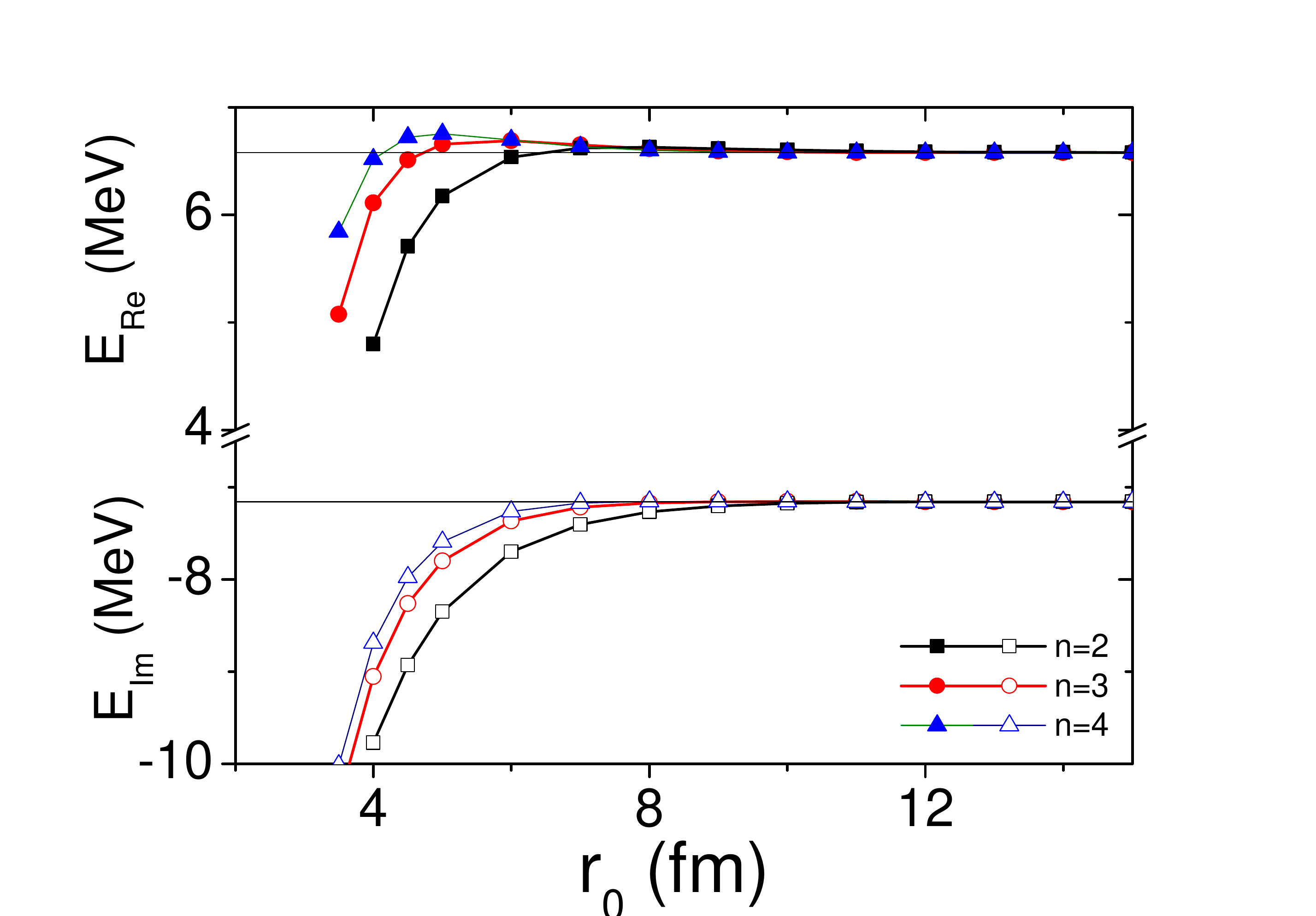}
\hspace{0.cm}
\centering\includegraphics[scale=0.3]{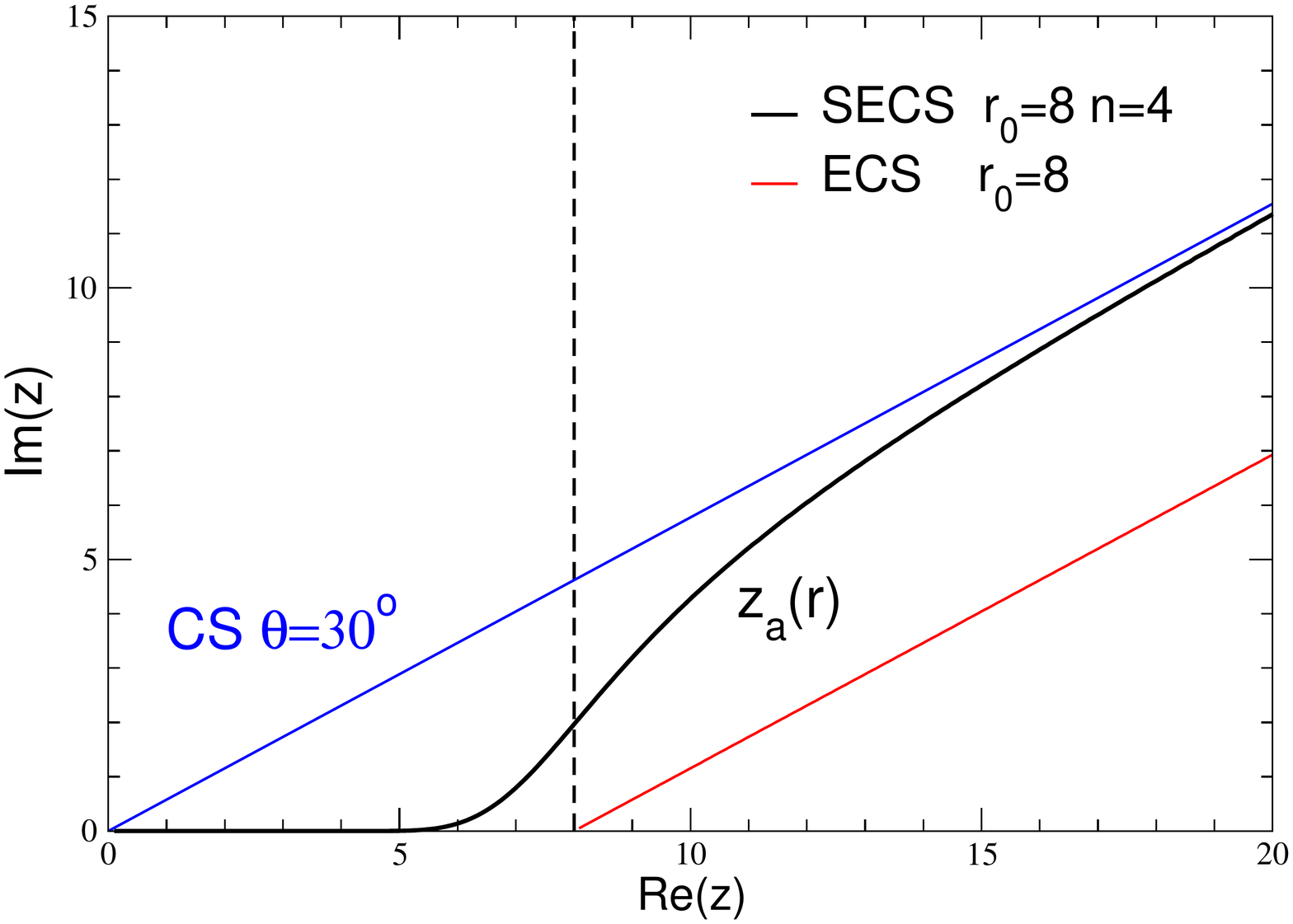}
\caption{P-wave resonance energies in the  2-body  potential
eq. (\ref{eq_espot_2b}) obtained by the SECS method with $n=4$ as a
function of  $r_0$ (left panel). If  $r_0$ is  larger  than the
interaction range  ($r_0\approx 8$ fm) one gets accurate
resonance position. The mapping  given in eq. (\ref{eq_sec_z}) used in these
calculations  for the converged value  is displayed in the right
panel.} \label{AECSM}
\end{figure}

To tackle this problem, the so called Exterior Complex Scaling
method (ECS)  was proposed \cite{ECSM_Simon_PLA71_1979}, which applies the CS
transform only in the region free from interaction, say $r\to
z(r)=r_0+re^{i\theta}$ with $r_0$ sufficiently large. However, in
its standard form, ECS has a formal inconsistency when applied to
the $A>2$ problem solved via  partial wave decomposition, due to
the interplay between different Jacobi coordinate sets. To
circumvent this problem we propose an analytic mapping $r\to
z_a(r)$ -- hereafter denoted SECS -- whose effect is extremely
small in the interaction region and, hence, it is applied to the kinetic energy term only.
If this mapping is performed by means of a sharp enough
function centered outside of interaction region, SECS gives very accuracy results.
In practice it can be realized by the means of a smooth step-like function $s(r)$:
\begin{equation}
r\rightarrow z_a(r)= s(r)re^{i\theta}+r [1-s(r) ]   \qquad
s(r)=e^{-(r_0/r)^n} \label{eq_sec_z}
\end{equation}
which is represented in Fig. \ref{AECSM} (right panel) together with its ancestors.

We  illustrate this method with  a two-body  P-wave resonance generated by the potential:
\begin{equation}
V(r)=V_r \frac{ e^{-\eta_r r} }{r}- V_a \frac{e^{-\eta_a r}}{r}
\label{eq_espot_2b}
\end{equation}
with $V_r$=2589.696 MeV, $V_a$=1128.393 MeV,  $\eta_r$=3.11
fm$^{-1}$,  $\eta_a$=1.55 fm$^{-1}$ and $\hbar^2/m$=41.471
MeV fm$^2$. In this example we took $\theta=40^\circ$, but the method applies well for larger angles.

We have plotted in Figure \ref{AECSM} (left panel) the complex
energy values of the resonance as a function of $r_0$, the range
parameter of SECS, calculated with the mapping (\ref{eq_sec_z})
and n=4 displayed in right panel. As one may see,  results  are
indistinguishable from the exact S-matrix pole
$E_R$+i$E_I$=6.578-i 7.1578   MeV for $r_0\gtrapprox$8 fm. With
$r_0=$12 fm, one reproduces four significant digits using  $n=3$
and $n=4$, while  the differences  using $n=2$ differs only by  1
keV. Despite the fact that our approach neglects the
transformation of the potential energy, it produces stable and
very accurate results.

When solving FY equations in $A\ge3$ sector, it is of interest to
apply SECS to hyperradius  $\rho$ only, for it  allows a consistent treatment of the kinetic energy operators in different Jacobi coordinate sets.
 We have successfully realized this option in 3-body calculations,
 but it is worth noticing that the transformation of four radial Jacobi coordinates turned to be even a better choice.
 The last procedure slightly breaks the consistency of the kinetic energy in the different FY components.
 Nevertheless when working with radial Jacobi coordinates we are able to evaluate the transformed kinetic energy
terms more accurately  if the transformation is natural  to these coordinates, which turns out to be a critical
ingredient in the calculations requiring sharp regulators.

\section{Results}\label{Results}

In this work  three different NN interactions have been considered: two realistic ones and a third one semi-realistic.
Among the first, the  I-N3LO chiral EFT potential of Entem and
Machleidt \cite{N3LO_EM_PRC68_2003} which was very successful in
describing the n-$^3$H elastic cross section near the resonance
peak even without corresponding three nucleon forces \cite{Bech_nt}, and the INOY
non-local potential from \cite{Doleschall_PRC69_2004} which
provides an accurate description of A=2,3,4 binding energies.
The semi-realistic S-wave MT13  potential \cite{MT_NPA127_1969} was also considered which, despite its bare
simplicity, describes well the gross properties of low energy few-nucleon systems. 

\begin{table}[h!]
\begin{center}
\begin{tabular}{ l l l l l l l l l }
           $J^\pi$                                      &    $2^-$   &                &    $1^-_1$   &                     & $0^-$         &                &      $1^-_2$                   &             \\\hline
                                                & $E_R$& $\Gamma$&  $E_R $    & $\Gamma $ & $E_R$      & $\Gamma$   &  $E_R$               &  $\Gamma$     \cr
I-N3LO                                       &   1.15(5)    &3.97(7)
& 0.90(3) & 3.80(8) &     0.78(15)        & 7.7(7)   &0.2(2) &
4.6(1) \cr INOY & 1.31(4) & 4.16(4) & 0.96(7)& 4.03(8) &0.80(15) &
8.3(1) & 0.2(2)&5.0(1) \cr
MT13 & 1.08(3) &4.06(5) &1.08(3) &  4.06(5)&1.08(3) &4.06(5) & 0.88(9) &4.4(1) \\\hline
RGM \cite{Arai_PRC66_2003}         &  1.52      &  4.11
&1.23        &   5.80 & 1.19 & 6.17  &    1.32 &  4.72  \cr tn
\cite{DGFJ_NPA786_2007} & 1.22    & 3.34          & 1.15 & 3.49
& 0.77 &  6.72 &1.15  & 6.38
\\\hline
Exp.  \cite{Tiley_A_4_NPA451_1992} &   3.19     & 5.42       &
3.50          & 6.73              &  5.27      & 8.92   &  6.02
& 12.99         \cr
\end{tabular}
\end{center}
\caption{Resonance parameters of $^4$H. Our results obtained with
three different potentials (first three rows) are compared to previous calculations using  a 2-body (t-n) potential \cite{DGFJ_NPA786_2007}
 or an RGM microscopic calculation \cite{Arai_PRC66_2003}.
Experimental R-matrix analysis \cite{Tiley_A_4_NPA451_1992} are  in the last line. All units are in MeV.}\label{Table_4H}
\end{table}

We have first computed the J$^{\pi}$=2$^-$,0$^-$,1$^-$ P-wave resonances of its closest isotope $^4$H, which
dominate the low energy n-$^3$H elastic cross section in the peak
region and which are intimately related with the resonant states in $^5$H. The method used here  is a direct
calculation of the 4N resonant state by solving the FY equations with the techniques developed in
\cite{LC_4n_PRC72_2005,16}   supplied with the corresponding asymptotically diverging boundary condition.
Results are displayed in Table \ref{Table_4H}, where all  values are given in MeV
and energies are relative to $^3$H threshold.
They are compared with previous calculations: the microscopic RGM 3-cluster model
 of \cite{Arai_PRC66_2003} and  the 2-body t-n model of \cite{DGFJ_NPA786_2007}.
As one can see, the three interactions we have considered in our
{\it ab initio}  calculations give close results, but are quite different from the  RGM   and
 t-n  model calculations, especially for the $1^-_2$ state which we predict as an almost subthreshold resonance  (third E-quadrant.
 Notice the degeneracy of the first  three states (2$^-$,1$^-_1$,0$^-$)  in the MT13 model. They all share total spin S=1 and total angular momentum L=1
without any coupling, due to the purely central and spin-spin terms of the interaction. The  1$^-_2$ state correspond to S=0.
For completeness, we also listed the values provided by an R-matrix analysis of experimental cross sections \cite{Tiley_A_4_NPA451_1992}.
Although for broad states only a qualitative comparison between S-matrix pole positions and R-matrix makes sense, we notice that in the present case the
resonance parameters can differ  by factors 2 to 5, what represents several MeV.

The  states of $^5$H nucleus are genuine 3-body final state
resonances, governed by a very complicated diverging behavior of
the wave function asymptotes. In this case we are not able to
directly impose the proper outgoing wave boundary conditions and
our calculations for this system rely on the ACCC and  the SECS
methods explained in the previous section.

\begin{figure}[h!]
\vspace{-.cm}
\begin{center}
\includegraphics[scale=0.3]{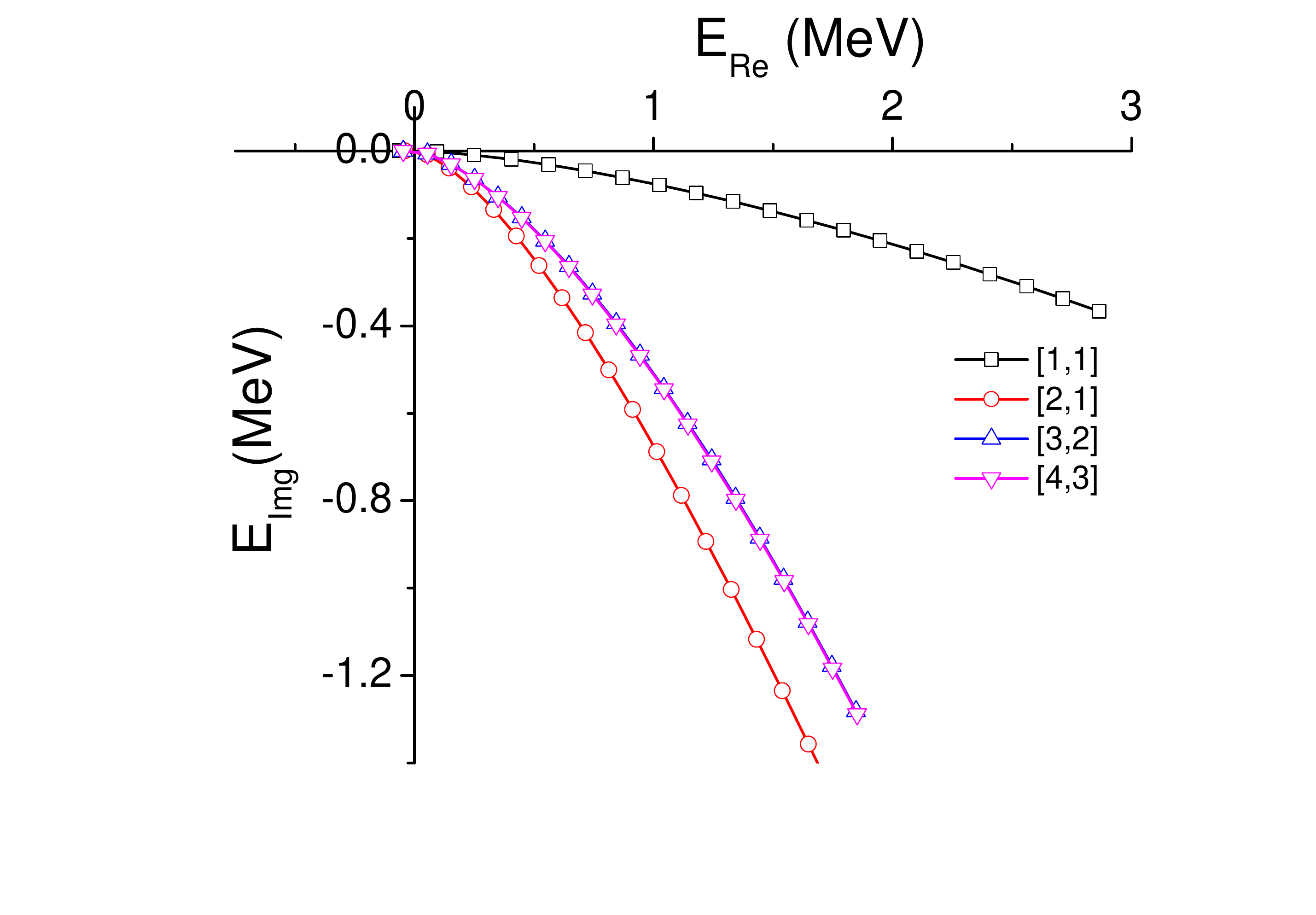}
\hspace{-1.cm}
\includegraphics[scale=0.3]{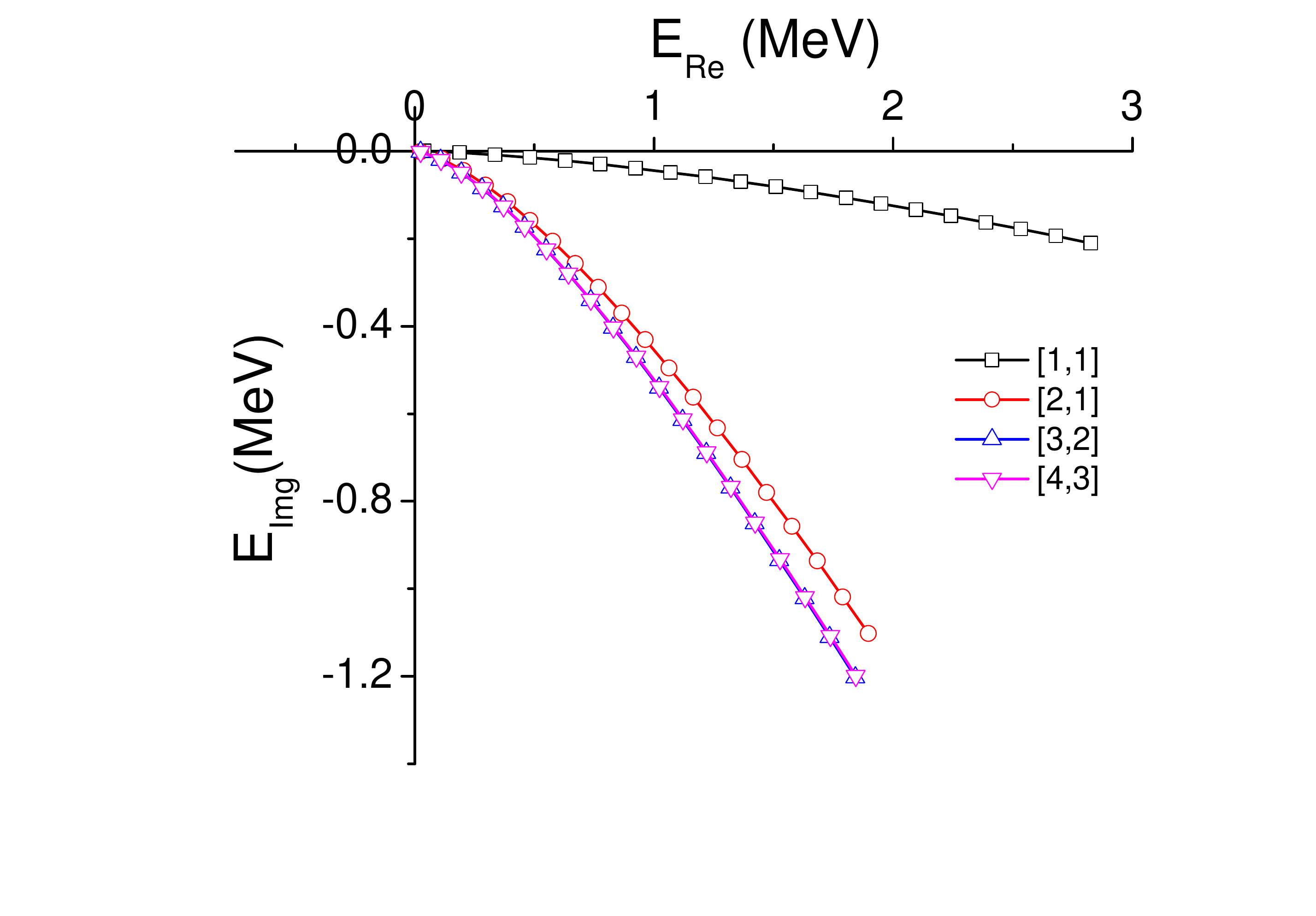}
\vspace{-1.cm} \caption{Resonance trajectories for J$^\pi$=1/2$^+$
state of $^5$H relative to $^3$H threshold. Each trajectory is
made by 20 equidistant values of the $V_5$  strength parameter
$\lambda$ in eq. (\ref{eq:accc_pot}), starting from nearly
threshold energy. The endpoint ($\lambda=0$) indicates the
extrapolated physical values of the resonance  predicted by the NN
interaction. Left and right  panels correspond to  different
parameterizations of $V_5$ ($\rho^2_0$=78.4 fm$^2$  and
$\rho^2_0$=108.9 fm$^2$  respectively, both with $p=0$). The
convergence of the results with respect to order of Pad\'{e}
approximant [m,n] are distinguished by colored curves.}
\label{fig:H5_trj_n3lo}
\end{center}
\end{figure}

We illustrate in Fig.~\ref{fig:H5_trj_n3lo} the ACCC results  for
the J$^\pi$=1/2$^+$ state with the  I-N3LO Hamiltonian. The
continuum extrapolation was  based on 23 binding energies,
starting from the triton threshold $B(^3H)=7.81$ MeV up to 300
MeV. Notice that this $B(^3H)$ value -- obtained in a 5-body
calculation with limited model space --  differs only by 0.04 MeV
from the fully converged 3-body calculations $B(^3H)=7.85$ MeV.
Though formally $m+n+1$ eigenergy entries suffice to determine
coeficients of Pad\'{e} approximant [m,n], as suggested in
ref.~\cite{ACCC_Kukulin_1989}, we have employed much larger
eigenenrgy entry sets by determining the coefficient using
least-square method. The last trick allows to minimize effect of
eigenergy set choice on the convergence of the Pad\'{e} series as
well as reduce the effect of numerical accuracy being finite. Two
different parameterizations of   $V_{5}$  potential
(\ref{eq:accc_pot}) were considered:  $\rho_0^2$=78.4 fm$^2$ and
$\rho_0^2$=108.9 fm$^2$, both with $p=0$.\footnote{In our previous
work \cite{Rimas_5N_FBS_2018}  parameterizations with different
values of $p$ were considered, providing results in agreement with
the $p=0$ case; however  its implementation requires a more
important numerical effort.} They correspond respectively to the
left- and right-panel of this figure. In both cases, the Pad\'e
extrapolation nicely converges to the resonance energy
E=1.85(1)-i1.25(5) MeV, a result which is  highly ensuring.
Nevertheless the convergence relative to  the order [m,n] of the
Pad\'e approximant should be considered with a grain of salt.
Indeed, one could expect that by increasing the order of Pad\'e
approximant, the accuracy in  the extrapolated energies should
improve accordingly. However this expected improvement saturates
at order [3,2], when the extrapolated energies are reproduced
within 100 keV. This numerical "saturation"  is certainly due to
the limited accuracy of the calculated 5-body energy inputs and
introduce  error bars in our results. In this context, increasing
the order of the  Pad\'e approximants makes no sense and may even
lead to numerical instabilities due to overfitting.

The extraction of  the same resonance parameters by means of SECS, provide  the value E=1.85(10)-i1.20(5), where errors
correspond to the variations of the rotation angle  $\theta\in$[30,45]$^\circ$ and the range parameter $r_0\in$[13,16] fm with n=3,4.

A similar study was performed for the INOY potential.
ACCC results,  based on $V_5$ parameters $\rho_0^2$=108.9 fm$^2$   were obtained.
A Pad\'e extrapolation was constructed from 28 B($^5$H) binding energies in the interval  [8.40,300] MeV.
They converge to the value E=1.73(5)-i1.19(9) MeV while SECS  provides this resonance at E=1.8(1)-i1.2(1) MeV.

\bigskip
Results for the J$^{\pi}$=1/2$^+$ ground state of $^5$H, with different interactions\footnote{MT13 results
with ACCC method were presented in \cite{Rimas_5N_FBS_2018}}
and the two different methods (ACCC and SECS) used to access complex energies, are summarized in Table 2.
As one can see,  both methods  provide resonance positions and widths compatible within the estimated errors.
 The width of this state turns out to be almost
independent of the employed interaction model -- including the
semi-realistic MT13 -- around a value
$\Gamma=2.4(2)$ MeV; its real part seems to be less stable. In
particular for MT13 potential real energy of the resonant state
seems to be slightly smaller than ones provided by the realistic
Hamiltonians. Apparently this might also be the reason for larger
deviation between the ACCC and SECS predictions -- one has
increasing difficulty to calculate accurately resonant states once
$E_i/E_R$ ratio increases. For J=3/2$^+$ state we were able to get
stable results only for MT13 potential and using SECS. As it was
the case for $^4$H, the J=3/2$^+$ and J=5/2$^+$ states resulting
from L=2 and total spin S=1/2 coupling, are degenerate due to the
absence of spin-orbit and tensor terms within this model.

\begin{table}[h!]
\begin{center}
\begin{tabular}{ l l l l l l l}
                                                    &    J=1/2$^+$   &                   &    J=3/2$^+$     &                      & J=5/2$^+$         &               \\\hline
                                                    &  $E_R$   & $\Gamma$&  $E_R $    & $\Gamma $  & $E_R$        & $\Gamma$ \cr
N3LO $^{(ACCC)}$                             &   1.8(1)    &
2.4(2) &                 &                      & & \cr
\phantom{N3LO} $^{(SECS)}$            &    1.9(2)    &  2.4(2) &
& & & \cr
INOY   $^{(ACCC)}$                           &     1.7(1)  &
2.4(2)       &                 &                      & & \cr
\phantom{INOY} $^{(SECS)}$            &     1.8(1) & 2.4(2) & & &
&             \cr
MT13    $^{(ACCC)}$                          &    1.4(1)  &
2.4(2) &                   &                        &        & \cr
\phantom{MT13} $^{(SECS)}$            &    1.7(2)  &   2.4(2) &
2.5(2)    &   3.8(4)          &     2.5(2)    &   3.8(4)
\\\hline
%
\cite{SDGB_PRC62_2000}    tnn     &  2.75(25)    & 3.5(5) & 6.6(3)
& 8 & 4.8(2) & 5 \cr \cite{DK_PRC63_2001} RGM     &    3.1(1) &
2.5(15) & & & & \cr \cite{Arai_PRC66_2003}        RGM     & 1.59 &
2.48 & 3.0 &      4.8                & 2.9 & 4.1   \cr
\cite{DGFJ_NPA786_2007} tnn     &   1.57        & 1.53 & 3.25 &
3.89 & 2.82          &   2.51 \cr \cite{ABV_JPCS_2008} RGM & 1.39
& 1.60 & 2.11 & 2.87            & 2.10 &   3.14  \cr
\cite{AD_NPA813_2008} RGM    & 1.9 (2)   &   0.6(2) & 4(1) &  3(1)
& &         \\\hline
Exp.  \cite{Korsheninnikov_PRL87_2001} & 1.7$\pm$0.3 &    1.9$\pm$0.4  &                  &                      &                       &          \cr
Exp.  \cite{Wuosmaa_PRC95_2017}        & 2.4$\pm$0.3  &  5.3$\pm$0.4 &                 &                      &                       &
\end{tabular}
\end{center}
\caption{Resonance parameters for the lowest $^5$H states. Our results, using different interactions and different
 methods to access the continuum, are in the first three rows.
They are compared to previous calculations using a t-n-n 3-body model \cite{SDGB_PRC62_2000,DGFJ_NPA786_2007}
or a microscopic RGM 3-cluster  \cite{DK_PRC63_2001,Arai_PRC66_2003,ABV_JPCS_2008,AD_NPA813_2008} description.
The most recent experimental findings are in the last rows \cite{Korsheninnikov_PRL87_2001,Wuosmaa_PRC95_2017}.
All units are in MeV.}\label{Table_5H}
\end{table}

Our results are compared with previous calculations. Refs
\cite{SDGB_PRC62_2000}   and  \cite{DGFJ_NPA786_2007}  are based
on a 3-body t-n-n approximation with an adjusted t-n two-body
potential and an ad-hoc 3-body force, while Refs.
\cite{DK_PRC63_2001}, \cite{Arai_PRC66_2003}, \cite{ABV_JPCS_2008}
and  \cite{AD_NPA813_2008} attempt a microscopic calculation
assuming a $^3$H+n+n cluster wave function, properly
antisymmetrized. Notice that these model calculations display
much larger fluctuations among them than our {\it ab initio} results with different potentials.

It is worth noticing that  the lowest resonant state of $^4$H is
significantly wider, i.e. has larger width, than the $^5$H one. This
suggest that the $^4$n system is less repulsive than the $^3$n one,
and
contributes in this way to stabilize the $^5$H isotope with
respect to $^4$H. If this $nn$ "pairing tendency" is pursued with
$6n$ one can indeed expect that $^7$H would be a narrower resonant
state than  $^5$H in agreement with GANIL's 2007 findings, where
a ground state of $^7$H is predicted with the astonishing resonance
parameters $E_R$=0.57  MeV and $\Gamma$=0.09 MeV. Conclusions from
the recent measurement performed at RIKEN with SAMURAI
multineutron detector are expected with interest.

For completeness we have also added in  Table 2,  the last two experimental results from
Refs. \cite{Korsheninnikov_PRL87_2001} and  \cite{Wuosmaa_PRC95_2017}.
The results  of  \cite{Korsheninnikov_PRL87_2001} are fully compatible with our calculations,
although, as we have already emphasized, any  direct comparison between experiments and theory
 is not very meaningful, apart from denoting the existence of the same S-matrix pole.

\begin{figure}[h!]
\centering\includegraphics[width=8.9cm]{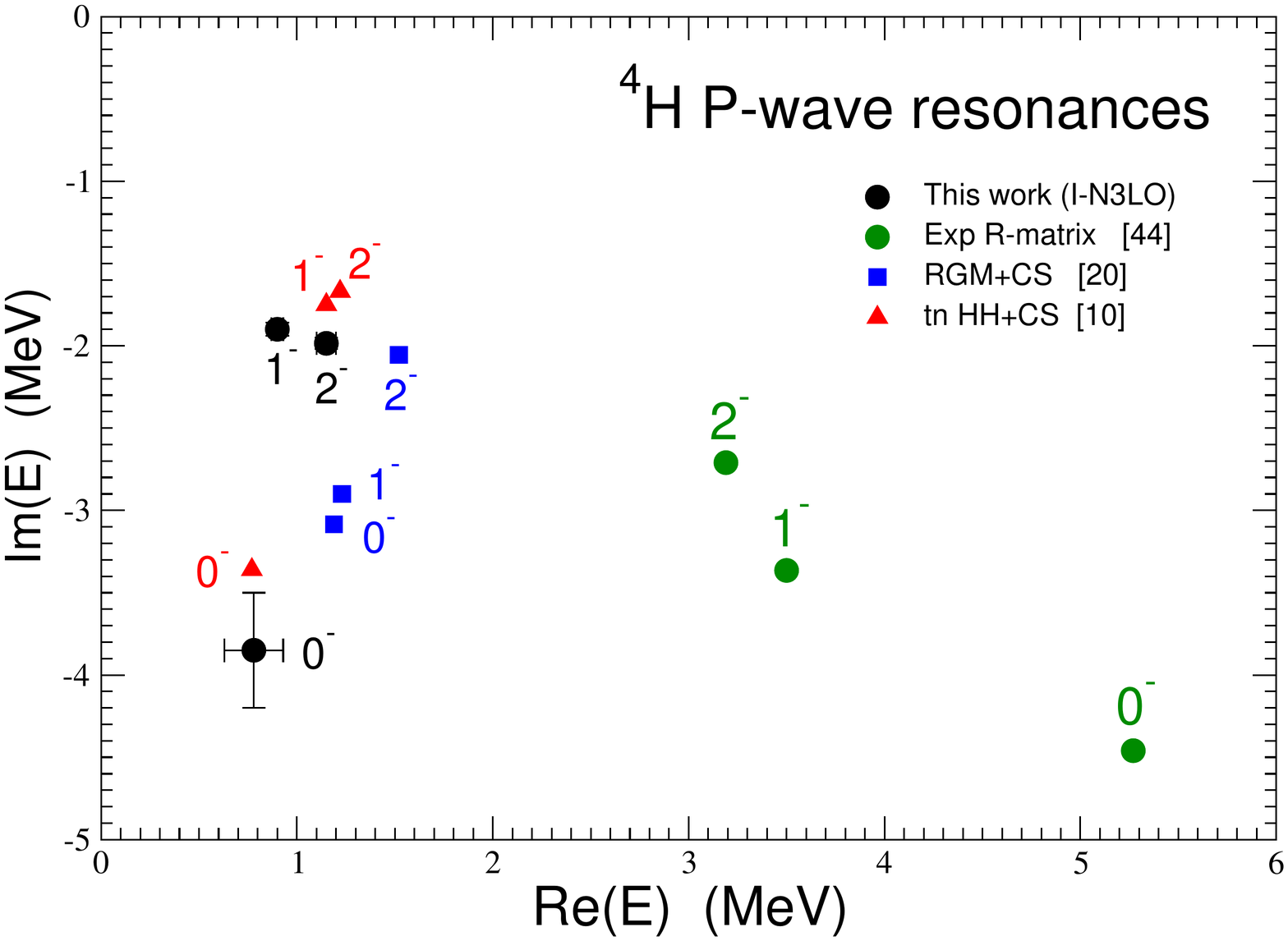}\hspace*{-1.1cm}
\centering\includegraphics[width=8.9cm]{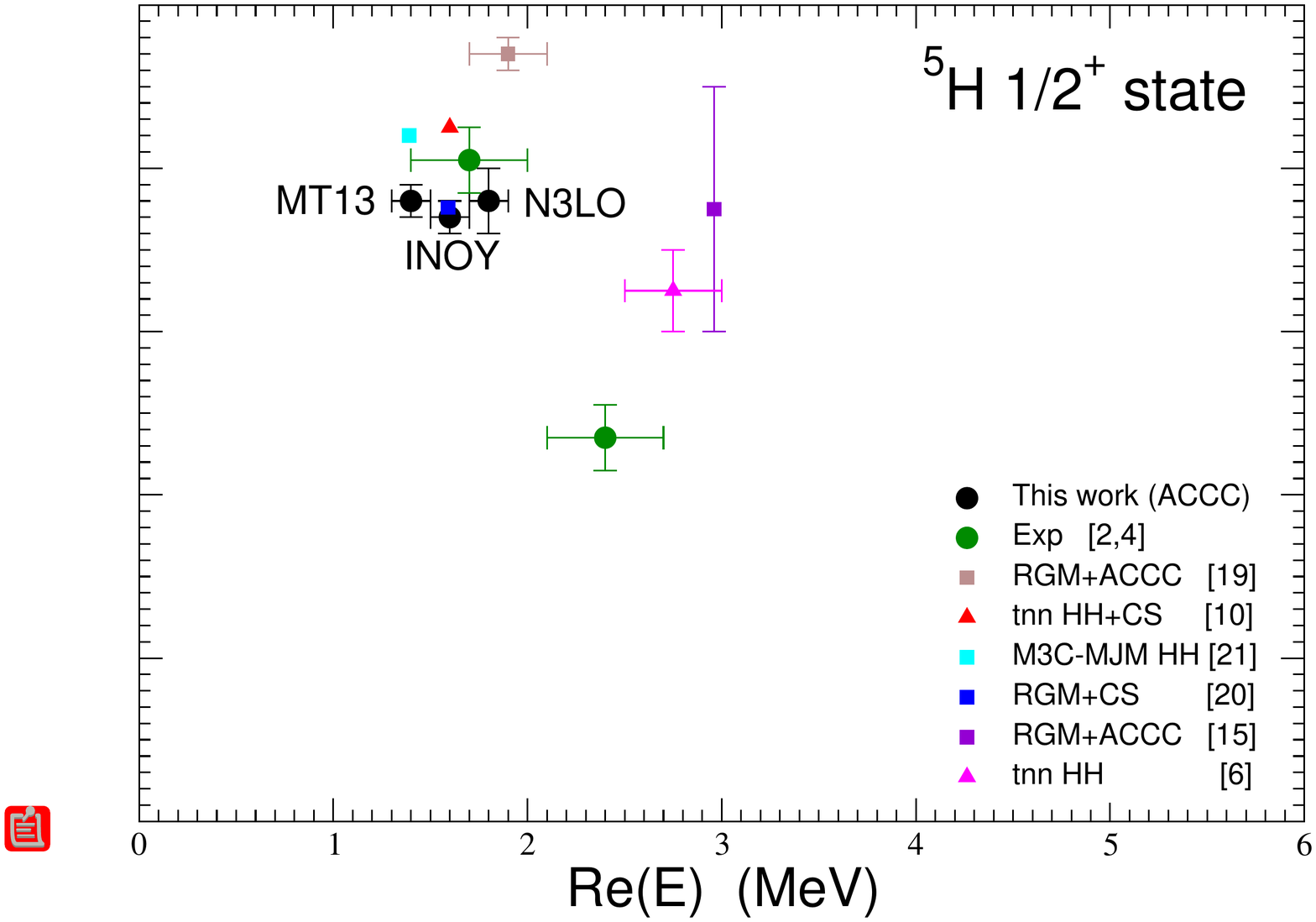}
\vspace{-.5cm}
\caption{Complex energy values of  the three lowest states of $^4$H (left panel) and to the  J$^{\pi}$=1/2$^+$ one of  $^5$H (right panel).
Energy scales in both figures are the same.
Our  results  are denoted by filled black dots and experimental values  in green ones.
Those  based on microscopic 3-cluster calculations  are denoted by squares and those based
on a 3-body  t-n-n  description by triangle up. Corresponding references are indicated in brackets.}\label{Fig_ALL}
\end{figure}

In order to allow the easiest  comparison of the ensemble of
results of Tables \ref{Table_4H}  and  \ref{Table_5H} we have
plotted  in Fig.~\ref{Fig_ALL}  the corresponding complex energy
values $E=E_R+i E_I$ (we recall that $E_I=-{\Gamma\over2}$). Left
panel corresponds to the three lowest $^4$H states and right panel
to the J=1/2  one in $^5$H. Our {\it ab initio}  results  are
denoted by filled black dots ($\bullet$). Those  based on a
microscopic 3-cluster calculations the same NN interaction are
denoted by colored squares and those based on a 3-body t-n-n
description by triangle up  symbols. The energy scales have been
taken the same in both figures to compare the proximity of $^5$H
and $^4$H resonances to the physical (real energy) axis and  their
potential impact on physical observables. The relative stability
of  $^5$H with respect to $^4$H is clearly manifested. Notice that
by comparing only the $E_R$ values of these two isotopes, that is
forgetting their imaginary part, we could have
reached an opposite conclusion. As one can see in this figure,
there is quite a large dispersion of the results, even among the
microscopic approaches. The positions extracted from the
experimental data
\cite{Tiley_A_4_NPA451_1992,Korsheninnikov_PRL87_2001,Wuosmaa_PRC95_2017}
are indicated by filled green circles. 

In the $^4$H case, the R-matrix analysis systematically predicts resonances parameters larger
than our theoretical values which,  as well recognized in ref.\cite{Tiley_A_4_NPA451_1992}, is quite a common
feature when dealing with broad resonant states. 
This trend is a clear consequence of  two approaches having different goals: the S-matrix approach tries to dtermine the analytical properties
of the underlying Hamiltonian, whereas the R-matrix analysis attempts to
describe the experimental observables with the smallest number of parameters. 
In the $^5$H case, since the resonant states turn to be 
narrower and more pronounced, one might expect a better agreement between the theoretically calculated S-matrix pole positions and the values
extracted from the experimental data. 
Indeed  the results  from \cite{Korsheninnikov_PRL87_2001} agrees well with our
calculated S-matrix poles, while those of \cite{Wuosmaa_PRC95_2017} provide significantly larger values - similar to the mismatch provided
by the R-matrix analysis in the $^4$H case.

\vspace{-.cm}

To conclude this section we would like to point out that the
methods we have used in this work to compute the resonant position
of $^5$H are strictly the same as those used in our previous
studies on the $3n$ and $4n$ systems
\cite{LC_3n_PRC72_2005,LC_4n_PRC72_2005,HLCK_PRC93_2016,CLHK_FBS_2017,LHC_PTEP7_2017}
where we refuted the existence of any experimentally meaningful resonant state.

\section{Conclusion}\label{Conclusion}

We have presented the first {\it ab initio} calculation to
determine the positions of $^5$H resonant states. This  has been
achieved by solving 5-nucleon Faddeev-Yakubovsky equations in
configuration space based on realistic nucleon-nucleon interaction
models.

Two independent methods were used to locate the resonance positions in the complex energy plane:
 a  variant of the smooth  exterior complex scaling method and the analytic continuation on the coupling constant.
The results show a remarkable stability with respect to the different tested interactions and support the recent experimental findings
\cite{Korsheninnikov_PRL87_2001,Wuosmaa_PRC95_2017},
although the direct comparison is biased due to the reaction mechanism involved in each experiment and to the indirect way the resonance parameters
are extracted.

The resonance parameters  of the J$^{\pi}$=2$^-$,0$^-$,1$^-$ states in $^4$H, which dominate the low-energy n-$^3$H
elastic cross section, have been also computed and  found to be slightly
wider than the ones for $^5$H ($\Gamma_{4H}\approx$ 4 MeV  for $\Gamma_{5H}\approx2.5$ MeV)   advocating  presence
of additional attraction of the 4n with respect to 3n system.
In view of that, any attempt to reproduce the $^7$H narrow state would be of the highest interest.

These {\it ab initio}  methods, recently applied to disprove the existence of any 3- and  4-neutron resonance
\cite{LC_3n_PRC72_2005,LC_4n_PRC72_2005,HLCK_PRC93_2016,CLHK_FBS_2017,LHC_PTEP7_2017}, are found
to be very efficient to compute the resonance parameters when they exist, as it is the case in $^4$H and $^5$H isotopes.

\section*{Aknowledgements}

Authors are indebted to  F. M. Marques for his careful reading of the manuscript and for useful discussions concerning the experimental results.
We were granted access to the HPC resources of TGCC/IDRIS under the allocation 2018-A0030506006 made by GENCI (Grand Equipement National de Calcul Intensif).
This work was supported by french IN2P3 for a theory project "Neutron-rich light unstable nuclei"
and by the japanese  Grant-in-Aid for Scientific Research on Innovative Areas (No.18H05407).



\begin{thebibliography}{10}

\bibitem{Young_PR173_1968}                  P.G. Young et al., Phys. Rev. 173 (1968) 949.
\bibitem{Korsheninnikov_PRL87_2001}   A.A. Korsheninnikov, et al., Phys. Rev. Lett. 87 (2001) 092501.
\bibitem{Golovkov_et_al_2003_04_05}    M.S. Golovkov et al,  Phys. Lett. B   566 (2003) 70, Phys. Rev. Lett. 93 (2004) 262501, Phys. Rev. C 72  (2005) 064612.
\bibitem{Wuosmaa_PRC95_2017}           A.H. Wuosmaa et al Phys. Rev. C 95   (2017) 014310.
\bibitem{H7_GANIL_2007}                       M. Caama\~{n}o et al., Phys. Rev. Lett. 99 (2007) 062502.


\bibitem{SDGB_PRC62_2000}                N.B. Shul'gina  B.V. Danilin, L.V. Grigorenko, M.V. Zhukov, J.M. Bang, Phys. Rev. C 62 (2000) 014312.
\bibitem{Grigorenko_nt_PRC_1999}        L.V. Grigorenko et al., Phys. Rev. C 57 (1998) R2099;  60  (1999) 044312.
\bibitem{GPT_nn_PLB32_1970}              D. Gogny, P. Pires, and R. de Tourreil, Phys. Lett.  B 32  (1970) 591.
\bibitem{GTZ_EPJA_2004}                      L.V. Grigorenko, N.K. Timofeyuk, and M.V. Zhukov, Eur. Phys. J. A 19,  (2004) 187.

\bibitem{DGFJ_NPA786_2007}               R. de Diego, E. Garrido, D.V. Fedorov, A.S. Jensen, Nucl. Phys. A 786 (2007) 71.
\bibitem{V_nn_GFJ_PRC69_2004}         E. Garrido, D.V. Fedorov, A.S. Jensen, Phys. Rev. C 69 (2004) 024002.

\bibitem{HOKY_NPA908_2013}               E. Hiyama, S. Ohnishi, M. Kamimura, Y. Yamamoto, Nucl. Phys. A 908 (2013) 29.
\bibitem{AV18_1995}                                R. B. Wiringa, V.G.J. Stoks and R. Schiavilla, Phys. Rev. C  51 (1995) 38.
\bibitem{SDEBZ_NPA597_1996}             N.B. Schul'gina, B.V. Danilin, V.D. Efros, J.S. Baagen, M.V. Zhukov, Nucl. Phys. A 597 (1996) 197.


\bibitem{DK_PRC63_2001}                      P. Descouvemont and A. Kharbach, Phys. Rev. C 63  (2001) 027001.
\bibitem{MN_VNN_TLY_NPA286_1977}  D. R. Thompson, M. LeMere, and Y. C. Tang, Nucl. Phys. A 286  (1977) 53.
\bibitem{MH_VNN_NPA459_1986}          T. Mertelmeier and H.M. Hofmann, Nucl. Phys. A 459 (1986) 387.
\bibitem{ACCC_Kukulin_1989}                 V.I. Kukulin, V.M. Krasnopolsky, J. Hor\`cek, Theory of Resonances, Kluwer, Dordrecht (1989).
\bibitem{AD_NPA813_2008}                     A. Adahchour, P. Descouvemont, Nucl. Phys. A 813 (2008) 252.

\bibitem{Arai_PRC66_2003}                     Koji Arai, Phys. Rev. C 68  (2003) 034303.

\bibitem{ABV_JPCS_2008}                      F. Arickx, J. Broeckhove, V. Vasilevsky, J. of Phys. Conf. Series 111 (2008) 012056.
\bibitem{Hasegawa_Nagata}                    A. Hasegawa and S. Nagata, Prog. Theor. Phys. 45 (1971) 786.
\bibitem{Rimas_5N_FBS_2018}              R. Lazauskas, Few-Body Syst. (2018) 59:13; doi.org/10.1007/s00601-018-1333-7.
\bibitem{Rimas_5N_PRC97_2018}         R. Lazauskas, Phys. Rev. C 97 (2018) 044002.

\bibitem{Miguel_RIKEN_NP1512}            F.M. Marqu\'es et al., RIBF Experimental Proposal NP1512-SAMURAI34.
\bibitem{Yakubovsky_SJNP5_1967}        O.A. Yakubovsky, Sov. J. Nucl. Phys. {5}  (1967) 937.
\bibitem{Fad_JETP39_1960}                    L.D. Faddeev, JETP 39 (1960) 1459.
\bibitem{Nogga_4a}                                  A. Nogga, H. Kamada, and W. Gloeckle Phys. Rev. Lett. 85 (2000) 944.
\bibitem{LazCarb_inoy}                            R. Lazauskas and J. Carbonell, Phys. Rev. C 70 (2004) 044002.
\bibitem{DeltFons_4N}                             A. Deltuva and A. C. Fonseca, Phys. Rev. C 76 (2007) 021001(R).

\bibitem{Baye_PR565_2015}                  D. Baye, Phys. Rep. 565  (2015) 1.

\bibitem{LC_3n_PRC72_2005}                R. Lazauskas and J. Carbonell, Phys. Rev. C { 71}  (2005) 044004.
\bibitem{LC_4n_PRC72_2005}                R. Lazauskas and J. Carbonell, Phys. Rev. C { 72} (2005)  034003.
\bibitem{HLCK_PRC93_2016}                 E. Hiyama, R. Lazauskas, J. Carbonell, N. Kamimura, Phys. Rev. C 93 (2016)  044004.
\bibitem{CLHK_FBS_2017}                      J. Carbonell, R. Lazauskas, E. Hiyama, M. Kamimura, Few-Body Syst. 58, 2  (2017) 58.
\bibitem{LHC_PTEP7_2017}                    R. Lazauskas, E. Hiyama, J. Carbonell, Prog. Theor. Exp. Phys. 7 (2017)  073D03.

\bibitem{CL_CS_SCAT_2011_13}            R. Lazauskas, J. Carbonell,  Phys. Rev. C 84 (2011) 044002;  Few-Body Syst. 54 (2013) 967.
\bibitem{ECSM_Simon_PLA71_1979}      B. Simon, Phys. Lett. A 71 (1979) 211.

\bibitem{N3LO_EM_PRC68_2003}          D.R. Entem and R. Machleidt, Phys. Rev. C 68 (2003)  041001.
\bibitem{Bech_nt}                                     M. Viviani, et al., Phys. Rev. C 84 (2011) 054010.
\bibitem{Doleschall_PRC69_2004}          P. Doleschall, Phys. Rev. C 69  (2004) 054001.
\bibitem{MT_NPA127_1969}                    R.A. Malfliet and J.A. Tjon, Nucl. Phys. A127 (1969) 161.

\bibitem{16}                                              R. Lazauskas  PhD, [http://tel.ccsd.cnrs.fr/documents/archives0/00/00/41/78/],  University J. Fourier, Grenoble, 2003.
\bibitem{Tiley_A_4_NPA451_1992}         D.R. Tilley, H.R. Weller, G.M. Hale, Nucl. Phys. A 541 (1992) 1.
\end{thebibliography}
\end{document}